\documentclass[aip, preprint]{revtex4-1}

\usepackage{graphicx}
\usepackage{dcolumn}
\usepackage{bm}
\usepackage{amssymb}
\usepackage{amsthm}
\usepackage{amsmath}

\usepackage[utf8]{inputenc}
\usepackage[T1]{fontenc}
\usepackage{mathptmx}
\usepackage{etoolbox}

\draft 
\usepackage{geometry}
\usepackage{url,hyperref,lineno,microtype,subcaption}
\usepackage{enumitem}
\newtheorem{theorem}{Theorem}

\newtheorem{proposition}{Proposition}

\newtheorem{lemma}{Lemma}
\newtheorem{remark}{Remark}

\makeatletter
\def\@email#1#2{%
 \endgroup
 \patchcmd{\titleblock@produce}
  {\frontmatter@RRAPformat}
  {\frontmatter@RRAPformat{\produce@RRAP{*#1\href{mailto:#2}{#2}}}\frontmatter@RRAPformat}
  {}{}
}%
\makeatother

\begin{document}


\title{Asymmetric Outcomes in Two-Actor Conflict Dynamics: Stability, Bifurcations, and Emergent Behaviors} 



\author{Eduardo Jacobo-Villegas}
\email{jac0bo@ciencias.unam.mx}
\affiliation{Universidad Nacional Autónoma de México, Av. Universidad 3000, C.U., Coyoacán, 04510, Ciudad de México, México}
\author{Josué Manik Nava-Sede\~{n}o}
\affiliation{Universidad Nacional Autónoma de México, Av. Universidad 3000, C.U., Coyoacán, 04510, Ciudad de México, México}
\author{Bibiana Obregón-Quintana}
\affiliation{Universidad Nacional Autónoma de México, 
Av. Universidad 3000, C.U., Coyoacán, 04510, Ciudad de México, México}

\date{\today}

\begin{abstract}
In this paper we present an analytical and numerical study of a generalized model of two-actor cooperative-competitive conflict of the Continuous Opinions and Discrete Actions (CODA) type. Theoretically, we note that the introduction of a new parameter allows generalizing feedback as strong and weak. Furthermore, we show that for positive-positive and negative-negative feedback there exists a supercritical pitchfork bifurcation, and that the model does not admit limit cycles in any case, and we study the effect of different parameter values and initial conditions by using a difference equation approximation of the model. 
Additionally, our model offers important insight on social phenomena such as false levels of support among cooperators, often observed in agreement negotiations; instances of ``non-strict consensus'' when two people support the same political position, albeit with different intensities; and competitive situations, such as in competitions with disproportionate profit and losses. Thus, this generalized model offers an enhanced descriptive power compared to previously proposed models.
\end{abstract}


\maketitle 

\begin{quotation}
Opinion dynamics is a major multidisciplinary area of study spanning psychology, sociology, economics, and political science. Its core objective is to investigate, model, and evaluate the evolutionary processes of actors' (or agents') opinions during social interactions in order to understand their emergence, persistence, variation, and dissemination. In mathematical modeling of opinion dynamics, actors' opinions are usually represented as numerical values and classified into discrete and continuous models according to the cardinality of the set containing all possible actors' opinions. 
Our study focuses on continuous models, specifically Continuous Opinions and Discrete Actions (CODA) models, where actors' opinions vary over a continuous range but display a discrete attitude or stance. In this context, mathematical models of conflict are invaluable not only for the rigorous study of complex human interactions, but also because they allow the simulation of situations and strategies without the need to observe or experiment in real-life settings that could prove unethical or dangerous, and they allow the identification of critical variables through easily controllable sensitivity analyses. 
In this article, we explore the analytical and practical consequences of generalizing a cooperative-competitive conflict model between two actors by incorporating a previously neglected parameter. Our findings reveal that the generalized formulation substantially improves the descriptive power of the original model.
\end{quotation}



\section{\label{Intro}Introduction}

Social conflicts between individuals, groups or nations  arise due to incompatibilities between their interests, goals or ethical values. Understanding their originating factors is of crucial importance, since they may offer a way to prevent, avoid, or regulate such conflicts, which may even escalate to international scales, affecting social cohesion, political stability, and economic development \cite{handbook, lewin1948resolving, burton1990conflict}.

The theory of cooperative and competitive conflict was developed in \onlinecite{Deutsch1949, Deutsch1973}. In these seminal works, it was suggested that the cooperative or competitive tendencies of the actors involved 
in a conflict are decisive to determine its development and outcome. In these works, a cooperative (respectively, competitive) relationship exists when the interdependence of the goals or values of the actors present a positive (respectively, negative) correlation. These ideas have been proven useful both in the development of psychological theories \cite{johnson2003social, Johnson21stCentury} and in practical applications, e.g., during experiments studying the advantages of competitive, cooperative, and interteam gamification in crowdsourcing systems \cite{coopcomp}, measuring and validating social interdependence during cooperative learning of health professionals \cite{ shimizu2020measuring}, assessing individual differences in competitive environments \cite{competitive} and determining the role of dominance in situations of cooperation and competition between adult siblings \cite{hernandez2023role}. 

Mathematical models of conflict allow the rigorous study of assumptions and conditions avoiding ambiguities, the simulation of situations and strategies without the need to observe or experiment in real settings which could prove unethical, the identification of critical variables through easily controlled sensitivity analyses, making predictions based on well-established assumptions, and the exact replication of previously reported studies.  

The inherent complexity in the interaction between multiple actors with heterogeneous interests, personalities, opinions, and influences has been a major challenge for the mathematical modeling of conflicts.
However, over the years, different approaches have been developed, such as evolutionary game theory models that analyze strategic decision-making \cite{madeo2020self,li2022game}, complex networks that study the structural-dynamic relationships in the formation and change of opinions \cite{larson2021networks, Lalo, Isakov_2019}, 
agent-based models simulating human societies to study the emergence of social phenomena \cite{lipiecki2022polarization,kan2023adaptive,alvarez2024mass}, and systems of nonlinear differential equations \cite{Liebo, sahasrabuddhe2021modelling} that allow us to gauge the effect of specific variables in conflictive environments.

In this paper, we use both analytical and numerical methods to study a generalization of the nonlinear ordinary differential equations model of cooperative-competitive conflict between two actors presented in \onlinecite{Liebo}, studying the effect of a previously disregarded parameter. This model, like the one presented in \onlinecite{Liebo}, considers the dynamical evolution of the state (or opinion) of each of the actors, taking into account their resistance to change and the feedback received from their neighbor.

The conflict model addressed here falls into the category of opinion dynamics models known as CODA, introduced in \onlinecite{martins2008continuous}, whose fundamental characteristic is considering that the choice of a specific attitude is the result of continuous and gradual changes in opinion. Unlike traditional CODA models, which use a Bayesian opinion updating approach \cite{martins2020discrete}, we do so using coupled differential equations, because in our analysis the opinions of the actors are dependent.

The main purpose of this work is to characterize novel scenarios resulting from the variation of the previously neglected parameter, in order to evaluate whether its inclusion expands the descriptive capacity of the original model. Furthermore, we also aim to obtain a complete analytic characterization of the model to determine the dynamical behavior of the individual actor states arising from specific initial conditions, which would provide better understanding for future research.

Our main analytical results consist of proving the existence of a pitchfork bifurcation when considering positive-positive (cooperation) and negative-negative (competition) feedback, as well as the nonexistence of limit cycles.
In addition, by analyzing both stable and unstable manifolds of the neutral equilibrium solution, as well as a difference equation approximation of the model, we obtain a full characterization of the temporal evolution of the system for specific initial conditions and parameter values.

We also find that this parameter explains previously observed behaviors including  counterintuitive increases in feedback strength, uneven steady states with equal signs, as well as with opposite signs but equal absolute values,
which we relate to social phenomena such as false or apparent levels of support among cooperators, situations of ``non-strict consensus'' and competitive systems with non-proportional profits and losses, respectively.

This paper is structured as follows. Section \ref{Model} introduces the model and outlines the techniques and parameters used in the analysis. Section \ref{Results} shows our model results, as well as a brief discussion. This section is divided into four subsections for clarity. The first subsection addresses weak feedback. We show that it is not essential to characterize the behavior in the positive-negative feedback case. In the next subsections, we examine strong feedback in the positive-positive and negative-negative feedback cases, while the last subsection analyzes the effects of the individual effect of each model parameter. Finally, in Section \ref{Conclusions} we discuss our results more broadly and outline directions for future research.

\section{\label{Model}Model and methods}

Our model is based off the nonlinear ordinary differential equation model of cooperative-competitive conflict between two actors introduced in \onlinecite{Liebo}, given by 
\begin{equation}\label{neg_2actors}
\begin{matrix}
dx_1/dt&=& -m_1x_1 + c_1\tanh(x_2)\\
dx_2/dt&=& -m_2x_2 + c_2\tanh(x_1),
\end{matrix}
\end{equation}
where $m_1, m_2>0$, $c_1, c_2\in\mathbb{R}$. Here  $x_i(t)$ represents the state (or opinion) of actor $x_i$ at time $t$, $m_i$ the bias of actor $x_i$ toward the neutral state 0, i.e. the resistance to change, and $|c_i\tanh(x_j)|$ is the feedback that the actor $x_j$ exerts on the actor $x_i$. 
 
 Although in Eq.~(\ref{neg_2actors}) the terms $m_1$ and $m_2$ are independent, in \onlinecite{Liebo} it was clearly stated that their study was restricted to the case $m_1=m_2$. The effects of time delays on this model were studied in Ref. \onlinecite{rojas2013time} for $m_1=m_2$. 

In this work, we study the most general version of Eq.~(\ref{neg_2actors}), given by 
\begin{equation}\label{NEW_2_actors_model}
\begin{matrix}
dx_1/dt&=& -m_1x_1 + c_1\tanh(p_1x_2)\\
dx_2/dt&=& -m_2x_2 + c_2\tanh(p_2x_1),
\end{matrix}
\end{equation}
where $x_i, m_1, m_2, c_1, c_2$ are as in the previous model (\ref{neg_2actors}) and we introduce the new parameters $p_1, p_2\in\mathbb{R}^+$, which can be interpreted as a measure of the threshold steepness in feedback that the actor $x_j$ exerts on the actor $x_i$, which is now given by  $|c_i\tanh(p_ix_j)|$. Note that these parameters are necessary for dimensional consistency of the model. Therefore, in Eq.~(\ref{neg_2actors}), these are implicitly set to one.
 
 Depending on $\operatorname{sgn}(c_i)$,  three types of feedback were considered for the systems (\ref{neg_2actors}) and (\ref{NEW_2_actors_model}): positive-positive feedback ($c_1, c_2 >0$); negative-negative feedback ($c_1, c_2 <0$); and positive-negative feedback ($c_1c_2<0$).  in \onlinecite{Liebo} it was determined that positive-positive and negative-negative feedback can be used to describe cooperative and competitive conditions, respectively, while positive-negative feedback was used to study the effect of a single actor on the long-term limit, by temporarily changing the type of feedback.  

Intuitively, the non-zero initial states of the actors can be considered as slight biases in opinion toward two possible positions, a positive and a negative one; and their long-term states as the position (or stance) resulting from a mutual feedback process, which can be cooperative, competitive, or cooperative-competitive. 

Nondimensionalizing the system of equations by using the following change of variables, $x(m_1 t)= p_2x_1(t)$, $y(m_1 t)= p_1 x_2(t)$, and $\tau=m_1 t$, Eqs.~(\ref{NEW_2_actors_model}) are transformed  into
\begin{equation}\label{Reduced_model}
\begin{matrix}
dx/d\tau&=& -x + p\tanh(y)\\
dy/d\tau&=& -qy + r\tanh(x),
\end{matrix}
\end{equation}
where $p= c_1p_2/m_1$, $r=c_2p_1/m_1$ and $q=m_2/m_1$. Note that $q$ is always positive and that the signs of $p$ and $r$ depend entirely on those of $c_1$ and $c_2$, respectively. Thus, we find that the problem has three degrees of freedom, while in \onlinecite{Liebo}, only changes in two parameters were explored.

Throughout this paper, we treat Eqs.~(\ref{NEW_2_actors_model}) and (\ref{Reduced_model})   interchangeably. We say that the system (\ref{Reduced_model}) has weak feedback if its parameters are such that $|pr|<q$ and strong feedback when $|pr|>q$.

As shown in the Appendix, we first compute analytically the equilibrium points of (\ref{NEW_2_actors_model}) for arbitrary values of  
all its parameters (see Propositions \ref{equilibrium_positive_and_negative_feedback} and \ref{0_onlyEquilibrium__mixed_feedback}), then we determine their stability by analyzing the eigenvalues of the linearized system (see Theorems  \ref{stability_positive_negative_feedback} and \ref{equilibrium_mix}). Finally we obtain a nonstandard
elementary stable finite-difference scheme for  (\ref{NEW_2_actors_model}), see Theorem \ref{Discretizacion_final_Modelo_OD}, since these schemes allow approximating systems of differential equations through systems of difference equations that are consistent with the original differential system, especially with the stability properties of the equilibrium points.

For all numerical results, random initial states in $(-1,1)$, and a step size $h = 1/50$ were used. These were obtained using code in Python. Phase portraits, were obtained using Wolfram. 
\section{\label{Results}Results and discussions}
\subsection{Weak feedback}
\begin{figure*}
\includegraphics[width=0.85\linewidth]{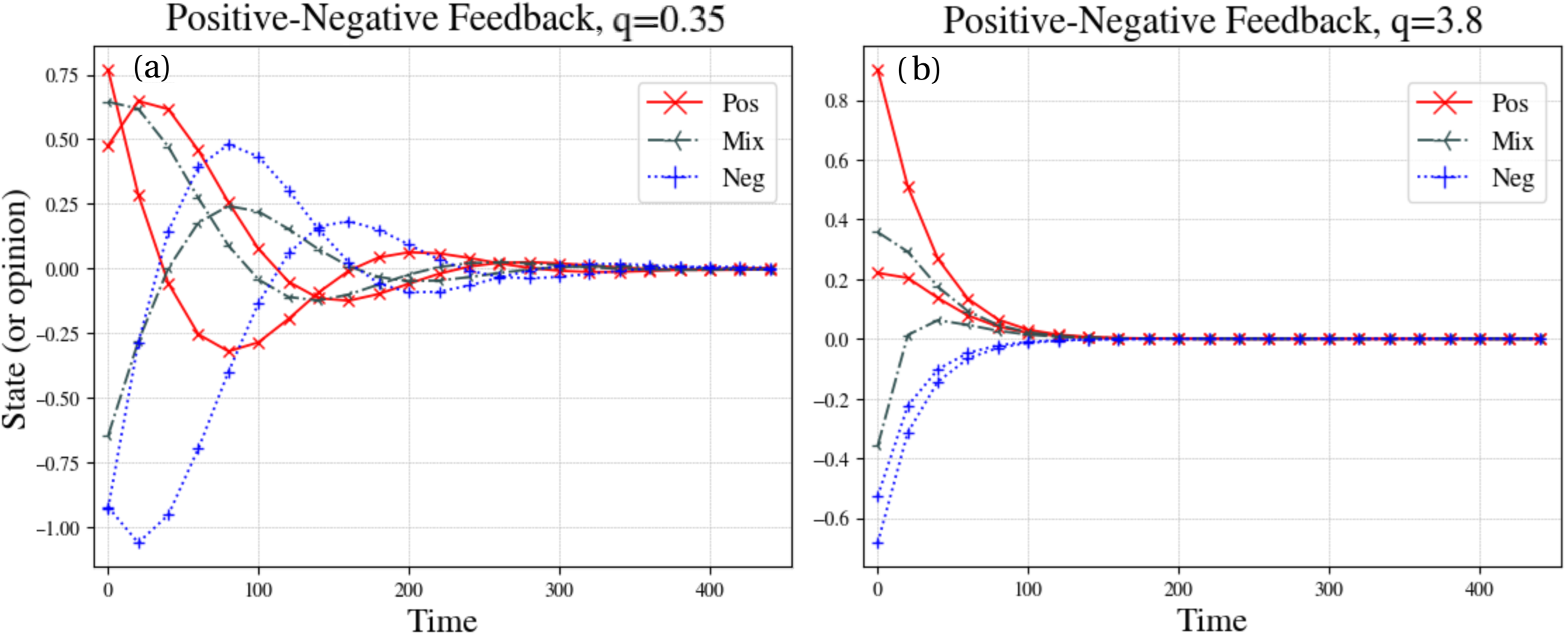}
\caption{\label{Figure_Pos_neg_evolutions}Time evolution of the opinions of the two actors in the system with positive-negative feedback. Three types of initial conditions are considered: both positive (red crosses), both negative (blue crosses) and with different signs (grey markers). The observed behavior agrees with Theorems \ref{0_onlyEquilibrium__mixed_feedback} and \ref{Limit_cycles}. In all cases, $r=-p=4/3$, so the threshold value was given by $4|pr|\approx 7.11$. (a) If $(q-1)^2<4|pr|$, the states oscillate towards equilibrium, (b) while if $(q-1)^2\geq 4|pr|$, the states decay monotonically towards the equilibrium.}
\end{figure*}

The weak feedback considered in this work (see Section \ref{Model}) generalizes the weak feedback defined in \onlinecite{Liebo}, where weak feedback was defined as $|c_i|< m_i$ and $p_i=1$, therefore, straightforwardly $|pr|<q$, which is the weak feedback in our generalization.\\
For the case of positive-negative feedback, both actors always evolve to a neutral state, since the long-term behavior does not depend on the relation between $q$ and $|pr|$. Theorems~\ref{0_onlyEquilibrium__mixed_feedback} and \ref{Limit_cycles} ensure that  this is the only possible behavior of the system in this case (Fig.~\ref{Figure_Bifurcation_diagram} (c) illustrates this fact).\\
In general, in the case of positive-negative feedback, it is true that the eigenvalues of the linearized system have nonzero imaginary part if and only if  $(q-1)^2<4|pr|$ (see the behavior of $\gamma$ in Lemma~\ref{Justificacion_hipotesisDiscretizacion2}); thus, $4|pr|$ is effectively a threshold which determines the type  of convergence towards the neutral state. If $(q-1)^2<4|pr|$, then $\hat{0}$ is a stable focus (see Theorem~\ref{0_onlyEquilibrium__mixed_feedback}), whereby the system oscillates towards the neutral state (Fig.~\ref{Figure_Pos_neg_evolutions} (a)), while for $(q-1)^2\geq 4|pr|$, no oscillations toward the neutral state are observed (Fig.~\ref{Figure_Pos_neg_evolutions} (b)). in \onlinecite{Liebo}, the presence of oscillating states was related to emotional fluctuations. Furthermore, since positive-negative feedback always brought the system to a neutral state, this was used to justify that an intractable negative-negative feedback conflict can be changed to a scenario with neutral outcomes for both actors by unilaterally changing the type of feedback exerted by one of the actors.\\
When feedback is weak and either positive-positive or negative-negative, then by Theorem \ref{stability_positive_negative_feedback} there is only one equilibrium point, $\hat{0}$, which is stable. Furthermore, Theorem~\ref{Limit_cycles} guarantees that the system has no limit cycles, so the system will always converge to the neutral state in the long time limit (see Fig.~\ref{Figure_Pos_pos_and_Neg_neg_evolutions} (c) and (f)). Thus, as in \onlinecite{Liebo}, actors are not influenced by each other when such influences are weaker than a threshold value defined in terms of their resistance to change (see the bifurcation diagrams in Fig.~\ref{Figure_Bifurcation_diagram} (a)-(b)). One interpretation could be that actors are too wary of changes in habits or thoughts, and such behavior may only be overcome if they feel too much social pressure \cite{veblen2017theory}. 
\subsection{Strong positive-positive feedback}
\begin{figure*}
\includegraphics[width=0.805\linewidth]{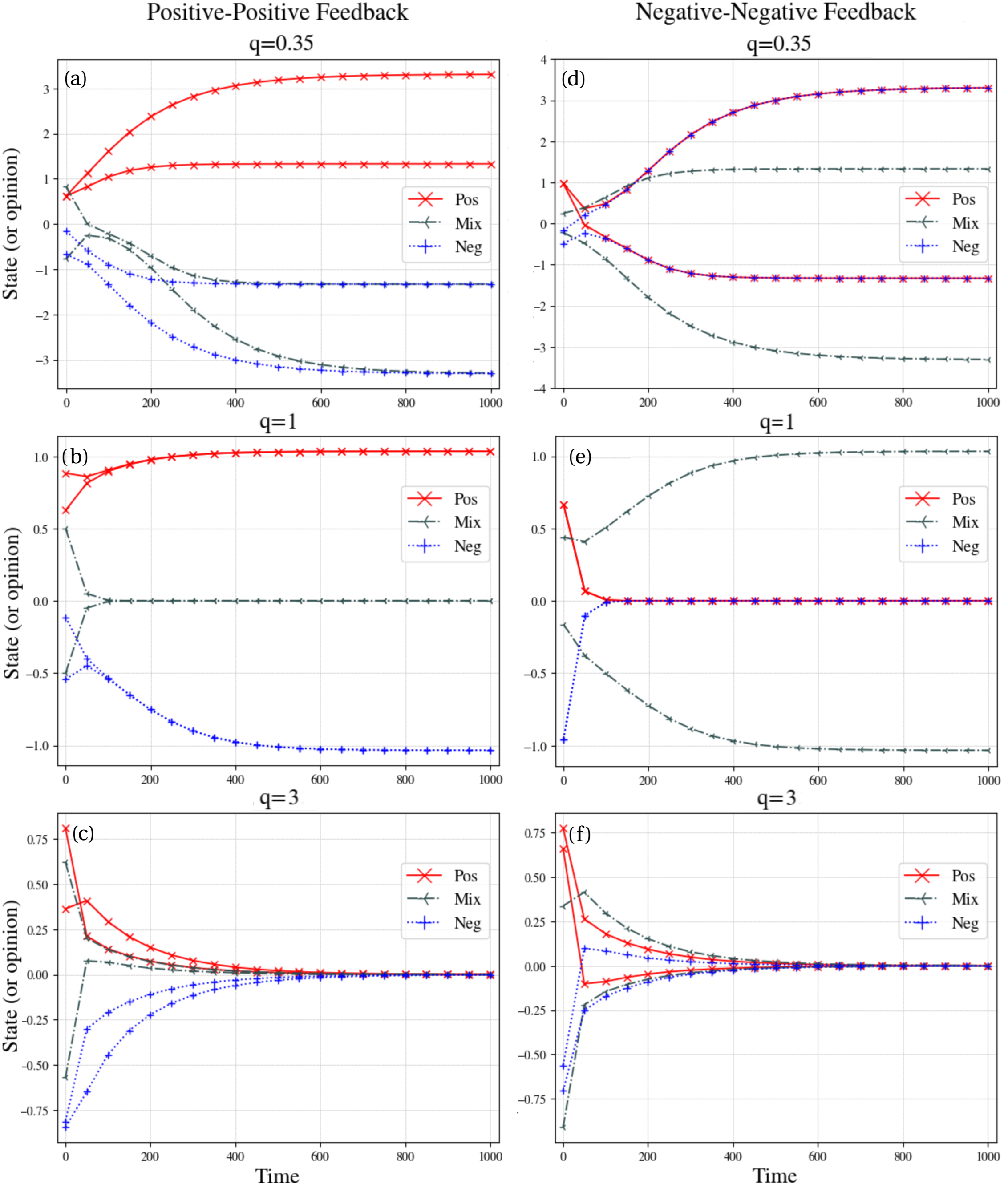}
\caption{\label{Figure_Pos_pos_and_Neg_neg_evolutions}Time evolution of the states, or opinions, of the two actors in the system with positive-positive and negative-negative feedback. Three types of initial conditions are considered: both positive (red crosses), both negative (blue crosses) and with different signs (grey markers). (a)-(c) show positive-positive feedback, (d)-(f) show negative-negative feedback. In all cases,  $|p|=|r|=4/3$, so the threshold value is $pr\approx 1.77$, thus if $q<pr$ (strong feedback), then the system evolves to one of the three existing equilibrium points (see (a)-(b) and (d)-(e)). On the other hand if $q\geq pr$ (weak feedback), then the system evolves to the only existing equilibrium point, $\hat{0}$ (see (c) and (f)).}
\end{figure*}

As with weak feedback, our definition of strong feedback (see Section \ref{Model}) generalizes strong feedback as defined in \onlinecite{Liebo}. In general, strong feedback corresponds to scenarios where the strength of influence experienced by each actor is stronger than their resistance to change (see Fig.~\ref{Figure_Bifurcation_diagram} (a)-(b)).\\
Whenever feedback is positive-positive (either strong or weak), we prove that, if both
actors' initial states have equal sign, then they never change sign (Theorem~\ref{Theorem_Discrete_statements} (a)). Furthermore, Theorems~\ref{stability_positive_negative_feedback} and \ref{Limit_cycles} ensure that there are three possible long-term behaviors, which we divide into two cases: 1) when $p=r$; and 2) when $p\neq r$.\\
In case 1), we have two subcases: $q=1$ y $q\neq 1$. If $q=1$, then the system is able to converge to the unstable equilibrium point, $\hat{0}$, if the actors have symmetric initial conditions about the horizontal axis (Fig.~\ref{Figure_Pos_pos_and_Neg_neg_evolutions} (b)), which follows from Theorem~\ref{manifolds}. Moreover, all other initial conditions lead
to exactly the same positive or negative value (as shown in Theorem~\ref{Theorem_Discrete_statements} c)). This behavior is related to what we call ``strict consensus'', when individuals manage to reach exactly the same position on an issue. This type of consensus has been considered in previous models of social influence \cite{kan2023adaptive} and consensus in information networks \cite{olfati2007consensus}, and could be relevant to social concepts such as unanimity in decision-making \cite{perez2018dynamic}.\\
On the other hand, if $q\neq 1$, then initial conditions almost everywhere veer the system away from the unstable equilibrium point, and both actors evolve towards positions with the same sign (positive or negative), but taking with unequal values (Fig.~\ref{Figure_Pos_pos_and_Neg_neg_evolutions} (a)). This behavior is related to ``non-strict consensus'', i.e., when the members of a group support the same position with different intensities, such as when two people support the same political stance, one of them moderately and another extremely so. In social psychology, it has been shown that the existence of a non-strict consensus is common, since factors such as goal interdependence \cite{Deutsch1973}, attitudinal strengths \cite{petty1998attitude} and latitudes of acceptance \cite{sherif1965attitude}.\\
The case 2) showed analogous behavior to the previous case; however, in this case “strict consensus” occurs for $q \neq 1$. 
Our findings suggest that ``strict consensus'' is indeed difficult to achieve and that, in general, cooperative environments generate ``non-strict consensus''. 
\subsection{Strong negative-negative feedback}
Whenever feedback is negative-negative (either weak or strong) and initially both actors' states have opposite signs, we show analytically that  their states evolve preserving their initial signs, i.e., they always have states with opposite signs (Theorem~\ref{Theorem_Discrete_statements} d)). Furthermore, if the system has strong negative-negative feedback, then analogously to the case of strong positive-positive feedback, we can consider three possible final configurations of the system (see Fig.~\ref{Figure_Bifurcation_diagram} (b)) within two cases: 1) $p=r$; and 2) $p\neq r$. \\
If $p=r$ and $q=1$, the system is able to converge to the unstable equilibrium point, $\hat{0}$, if the actors have exactly the same initial values (Fig.~\ref{Figure_Pos_pos_and_Neg_neg_evolutions} (e)),  as proven in
 Theorem \ref{manifolds}. Interestingly, this implies that if two competitive individuals have exactly the same opinion on a topic initially, they will end up with a neutral opinion on the matter, i.e., the conflict converges to a regulated situation. Also, we observe that if the initial values of the two actors are not equal, then their states evolve toward different but symmetric positions (as shown in Theorem~\ref{Theorem_Discrete_statements} f)), which could model situations we term ``strict polarization'', occurring when two actors support opposing positions with exactly the same intensity. This situation has been considered in several mathematical models as zero-sum games \cite{von2007theory}, Persuasive Arguments Theory \cite{barrera2020polarizing} and modified versions of Hegselmann-Krause type models \cite{anteneodo2017symmetry}. \\ 
On the other hand, if $q\neq 1$, the system almost surely never evolves to the unstable equilibrium point, since the set of initial conditions leading to a convergence toward the neutral state (its stable manifold) does not have a closed form (see Theorem \ref{manifolds}). Thus, the system evolves toward states with different signs that are not symmetric about the horizontal axis with almost every initial condition (Fig.~\ref{Figure_Pos_pos_and_Neg_neg_evolutions} (d)). 
This condition could model competitive situations in which two individuals support opposing positions with different intensity, for example, when supporting opposing political stances, with and without radicalization. Other research studies have addressed the role of profit asymmetry among:  competitors with kinship relationships \cite{granroth2013asymmetry}, hypercompetitive attitudes \cite{ryckman1990construction}, and aversion to losing in tournaments \cite{eisenkopf2013envy}.\\
If $p\neq r$, the system showed analogous behavior to the previous case; however, in this case ``strict polarization'' occurs for $q\neq 1$.  \\
Our analysis on strong negative-negative feedback suggests that competitiveness can lead to a regulated state (i.e., the neutral state) only under extraordinary conditions, as a clear ``winner'' and ``loser'' emerges from almost every initial condition.

\subsection{The individual effect of $\mathbf{m_i, c_i}$ and $\mathbf{p_i}$ on strong positive-positive feedback}
\begin{figure*}
\centering
\includegraphics[width=0.83\linewidth]{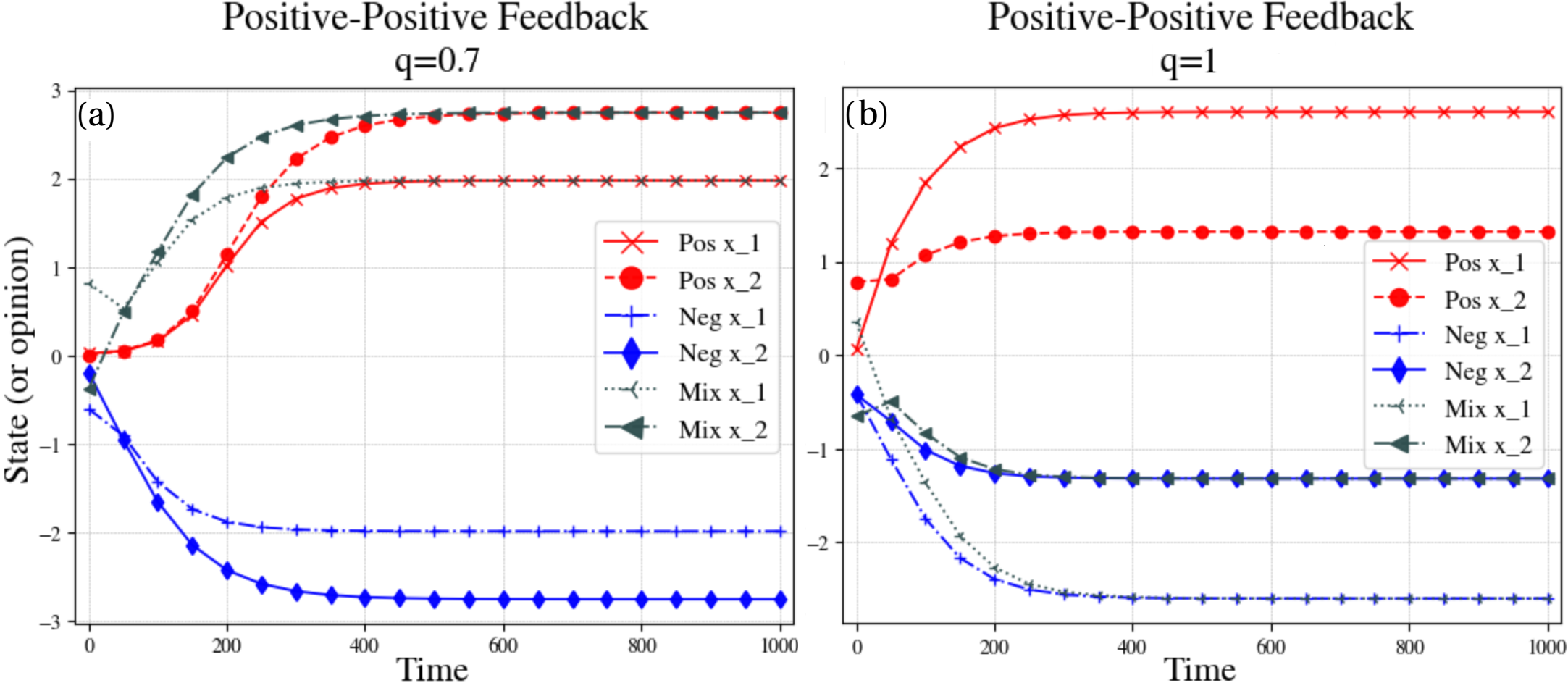}
\caption{\label{Figure_mi_pi_parameters}Individual effect of parameters $m_i$ and $p_i$ on systems with strong positive-positive feedback. Three types of initial conditions are considered: both positive (red crosses), both negative (blue crosses), and with opposite signs (grey markers). In (a) $m_1=1, m_2=0.7$ and $p=r=2$, $x_1$ converges to a final state closer to its initial state than actor $x_2$. In (b) $m_1= m_2 =1$, $c_1=c_2=3/2$, $p_1= 8/9$ and $p_2 = 2$, $x_1$ converges to a value farther away from its initial state than actor $x_2$.}
\end{figure*}

In this section we study the individual effects of the parameters $m_i, c_i$ and $p_i$ in the steady state under strong positive-positive feedback.\\
When $p=r>1$ and $q=(m_2/m_1)<1$ in Eq.~(\ref{Reduced_model}), actor $x_1$ converges to a final state closer to its initial state than actor $x_2$ (Fig.~\ref{Figure_mi_pi_parameters} (a)). On the other hand, when $q>1$, the situation is reversed. Intuitively, $m_i$ represents the resistance to change of $x_i$, so the aforementioned behavior is not unexpected.  Note that $p>1$ is crucial since otherwise, no change in the relative distance of the final states of the actors, with respect to their initial states, would be observed for different values of $q$ (see Remark~\ref{Remark_P_condition}).\\
Furthermore, it should be noted that $p=r$ implies that a behavioral change occurs for $q=1$; otherwise such a change would occur for $q\neq 1$, because the neighboring feedback forces would be unbalanced.\\
Furthermore, one should consider the case when all parameters other than $p_i$ are equal in Eq.~(\ref{NEW_2_actors_model}) and $p>1$, in a similar vein to our previous analysis. When $p_1<p_2$, counterintuitively actor $x_1$ converges to a state farther away from its initial state than actor $x_2$ (Fig.~\ref{Figure_mi_pi_parameters} (b)), and vice versa.. This is unexpected because $p_i<p_j$ implies that actor $x_j$ receives greater feedback, so it could be thought that it should converge to a final state farther away from its initial state than $x_i$. However, considering the transformed system, Eq.~(\ref{Reduced_model}), it can be observed that $x_1$ depends on $p$ and $x_2$ depends on $r$, such that $p_1<p_2$ implies $r<p$. Thus, the net effect of the individual parameters $p_1, p_2$ on the model given by Eq.~(\ref{NEW_2_actors_model}), is determined indirectly through parameters $p, r$ of the transformed model, Eq.~(\ref{Reduced_model}).\\
The apparently counterintuitive effect of $p_i$ can be interpreted as a situation with false support in cooperative environments; for example, when two individuals support the same position, but one of them is resentful (``I am jealous, but in a good way!''). It could also relate to environments where people with opposing interests momentarily decide to cooperate out of convenience, pressure or simply to avoid confrontation, as in agreement negotiations \cite{fisher2011getting}.\\
Finally, when all parameters other than $c_i$ are equal in the system, Eq.~(\ref{NEW_2_actors_model}), and $p>1$, then the parameters $c_i$ are directly related to the final system configurations, i.e., if $c_i<c_j$, then $x_j$ converges to a final state farther away from its initial state than $x_i$. This is expected since actor $x_j$ receives the strongest feedback. This fact reveals an essential difference between $c_i$ and $p_i$: while the effect of $c_i$ is intuitive in the final system configurations, that of $p_i$ is not.

\section{\label{Conclusions}Conclusions}
In this work, we exhaustively characterized a two-actor conflict model obtained by considering every degree of freedom in a model first introduced in \onlinecite{Liebo}. Such a broad characterization naturally led to a generalization of some previously defined concepts.\\
On a dimensional basis, we introduced two previously undefined parameters, $p_1$ and $p_2$, and non-dimensionalized the system to conclude that the system has three degrees of freedom, corresponding to the non-dimensional parameters $p$, $q$ and $r$. It should be noted that, in \onlinecite{Liebo}, the system was not fully characterized, since $m_1=m_2$ (i.e. $q=1$) and $p_1=p_2=1$.  Using the non-dimensionalized system, we characterized equilibrium points, their stability, bifurcations, and the basins of attraction of the stable steady states.\\
Both our newly introduced parameters, $p_i$, and the original model's parameters, $c_i$, are related to the strength of feedback between actors. However, we saw that their effects on both actors' states can be complementary, as well as unexpected at first glance.\\
The analysis of strong positive-positive feedback suggests that ``strict consensus'' is, in practice, impossible to achieve and that, in general, cooperative environments generate ``non-strict consensus''. On the other hand, our analysis considering strong negative-negative feedback suggests that a competitive relationship between two people can converge to a regulated state only under artificial conditions, so a clear ``winner'' and ``loser'' is expected. These results were analytically demonstrated by the existence of a pitchfork  bifurcation.\\
We were able to determine dynamic behavior depending on specific initial conditions. For example, the sign of the states of the actors remains invariant over time if the initial states of both actors have the same sign regarding positive-positive feedback, and if they have opposite signs  considering negative-negative feedback.\\
We found that our more general and dimensionally correct formulation of the model explains three situations which were previously unaccounted for, namely, cooperation with false or uneven levels of support; cooperation leading to ``non-strict consensus'';  and competitions with disproportionate gains and losses.\\
In upcoming publications, we will build upon these results by considering more general and realistic conflict scenarios. Specifically, we will model conflict among more than two actors, taking into account different relations and feedback among them, including but not limited to, higher-order relations between actors. Furthermore, we will study the effect of sudden disruptions and the return probability to the steady states with noisy communication, using a stochastic extension of the current model. 
\begin{acknowledgments}
We thank L.  Guzmán-Vargas for useful suggestions. This work was partially supported by UNAM-DGAPA (PAPIIT-IN114724) and SECIHTI (México).
\end{acknowledgments}
\section*{AUTHOR DECLARATIONS}
\subsection*{Conflict of Interest}
The authors have no conflicts to disclose.
\subsection*{Author Contributions}
\noindent \textbf{Eduardo Jacobo-Villegas:} Conceptualization (equal); Methodology (equal); Formal analysis (equal); Investigation (equal); Writing – original draft (lead); Visualization (lead). 
\textbf{Josué Manik Nava-Sede\~{n}o:}
Conceptualization (equal); Methodology (equal); Validation (lead); Formal analysis (equal); Investigation (equal); Resources (equal); Writing - review \& editing (lead). Supervision (equal).
\textbf{Bibiana Obregón-Quintana:} Conceptualization (equal); Methodology (equal); Formal analysis (equal), Resources (equal); Writing - review \& editing (supporting); Supervision (equal).
\section*{Data Availability Statement}
Data sharing is not applicable to this article as no new data were created or analyzed in this study.

\appendix
\section{Equilibrium points and their stability}
\begin{figure*}
\includegraphics[width=0.85\linewidth]{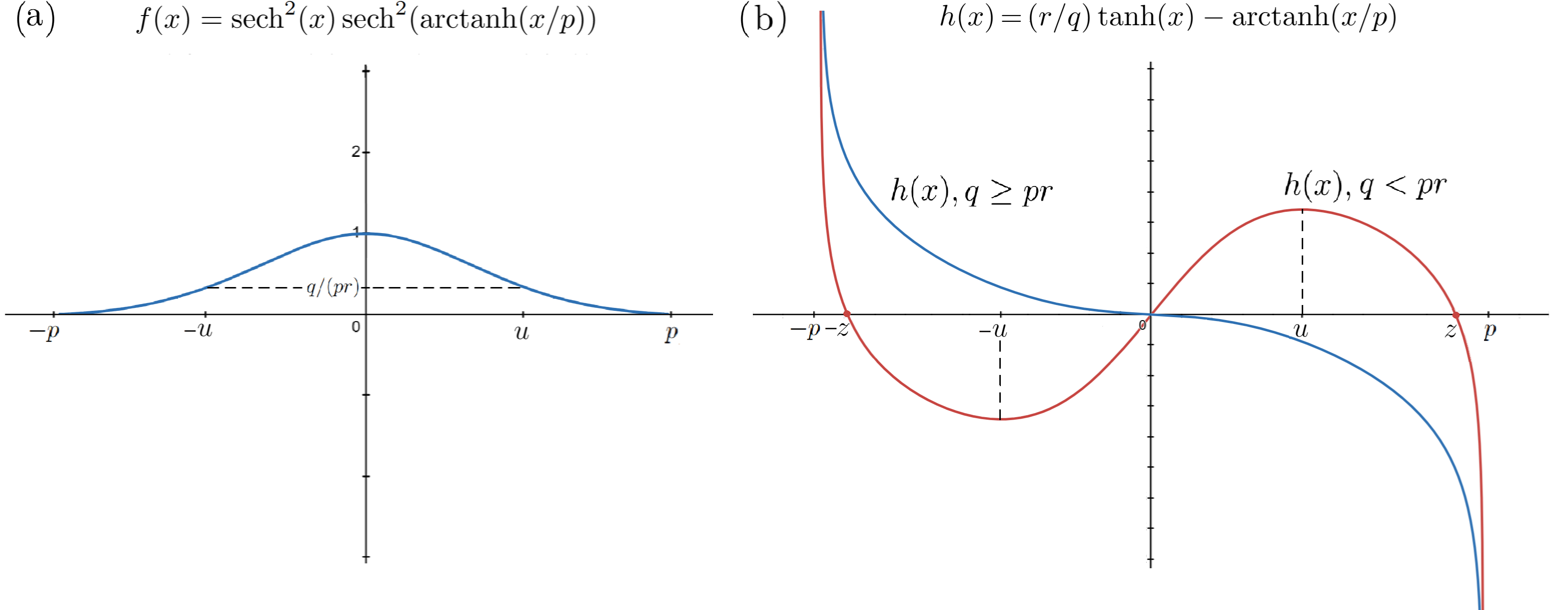}
\caption{\label{h_and_f}Functions related to Lemma \ref{Lemma_h_function}. (a) A sketch of the graph of the function $f(x)= \operatorname{sech}^2(x) \operatorname{sech}^2(\operatorname{arctanh}(x/p)), x \in (-p, p)$, if $q<pr$, there exists $u\in (0,p)$ such that $f(u)=q/(pr)$. (b) A sketch of the graph of the function $h(x)=(r/q)\tanh(x)-\operatorname{arctanh}(x/p), x \in (-p, p)$, when $q<pr$ (red) and when $q\geq pr$ (blue).}
\end{figure*}

\begin{lemma}\label{Lemma_h_function}
Let $q, p, r$ as in (\ref{Reduced_model}) and 
\begin{center}
$h(x) = (r/q)\tanh(x)-\operatorname{arctanh}(x/p),\quad x\in (-p, p)$,
\end{center}
if $p,r>0$, then 
\begin{itemize}[leftmargin=0.7cm, itemsep = 0.05cm, topsep=0.05cm]
    \item[i)] 0 is the only root of $h$, if $q\geq pr$,
    \item[ii)] there exists $z\in\mathbb{R}^+$ such that $z,-z$ and $0$ are all the different roots of 
 $h$, if $q< pr$.
\end{itemize}
\end{lemma}
\begin{proof}
Clearly $0$ is a root of $h$. Since $\tanh$ is an odd function, $\operatorname{arctanh}$ is also an odd function, it follows that $h$ is an odd function, therefore the proof reduces to showing that $h$ has exactly one root in $(0,p)$ if $q< pr$, and none if $q\geq pr$. 

Straightforwardly,
\scalebox{0.85}{$h'(x) = (r/q)\operatorname{sech}^2(x)-\dfrac{1}{p \operatorname{sech}^2(\operatorname{arctanh}(x/p))}$},
thus,
\begin{eqnarray}\label{h_critic_points}
\small h'(x) =0 \iff q/(pr) &=& \operatorname{sech}^2(\operatorname{arctanh}(x/p))\nonumber\\
\small &&\times \operatorname{sech}^2(x).
\end{eqnarray}
Differentiating again,
\begin{eqnarray}\label{2_derivate_h}
h''(x) &=& -(2x[p^3 \operatorname{sech}^4(\operatorname{arctanh}(x/p))]^{-1}\nonumber\\ &&+ 2(r/q)\operatorname{sech}^2(x)\tanh(x)).
\end{eqnarray}
Consider
    $f(x)= \operatorname{sech}^2(x) \operatorname{sech}^2(\operatorname{arctanh}(x/p)), x \in (-p, p)$.
We have that $f(x)>0, x \in (-p, p)$. Because $\operatorname{arctanh}$ is an odd and increasing function, it follows that 
\begin{equation}\label{f_even_decreasing}
f \text{ is even and decreasing on } [0,p).
\end{equation}
Since $f(0)=1$ and $\lim_{x\rightarrow p}f(x)=0$, it follows that $  f[(-p,p)] = (0,1]$, 
the graph of $f$ is shown in Fig.~\ref{h_and_f} (a).

\textbf{Case 1:} $q< pr$.
In this case, $q/(pr)<1$, then since $f[(-p,p)] = (0,1]$, it follows that there exists a unique $u\in (0,p)$ such that $f(u)=q/(pr)$; then by (\ref{h_critic_points}) it follows that $u$ is the unique critical point of $h$ in $(0,p)$.
Using (\ref{2_derivate_h}), $u$ is a local maximum of $h$. Note that $h(u)>0$, because otherwise there would exist an intermediate value $z_0\in (u-\epsilon,u)$, $\epsilon>0$ such that $f(z_0)>f(u)$, which is impossible. Since 
$\lim_{x\rightarrow p}h(x)=-\infty$, 
then the Mean Value Theorem allows us to state that there exists $z\in (u,p)$ such that $h(z)=0$. Moreover, $z$ is the only number with this property in $(0,p)$, otherwise by Rolle's Theorem, there would exist $z_1\in (0,p)$ such that $z_1\neq u$ and $h'(z_1)=0$, which is impossible because $u$ is the only critical point of $h$ in $(0,p)$. The red line in Fig.~\ref{h_and_f} (b) shows a sketch of the graph of $h$ in this case.

\textbf{Case 2:} $q\geq pr$.
In this case, $q/(pr)\geq 1$. Since $f(0)=1$, then by (\ref{h_critic_points}) and (\ref{f_even_decreasing}), it follows that $h$ does not have critical points in $(0,p)$. Note that this implies that $h$ has no root at $(0,p)$, since if there were $z_3\in (0,p)$ such that $h(z_3)=0$, then by Rolle's Theorem there would be a critical point of $h$ at $(0,z_3)$, which is impossible. The blue line in Fig.~\ref{h_and_f} (b) shows a sketch of the graph of $h$ in this case.   
\end{proof}

\begin{proposition}\label{equilibrium_positive_and_negative_feedback}
If the system (\ref{Reduced_model}) is considered with either positive-positive or negative-negative feedback, i.e. $pr >0$, then
\begin{itemize}[leftmargin=0.7cm, itemsep = 0.05cm, topsep=0.05cm]
    \item[i)] $\hat{0}$ is the only equilibrium point, if $q\geq pr$,   
    \item[ii)] there are exactly three different equilibrium points, if $q< pr$, given by $\hat{0}$, $(z,w)$ and $-(z,w)$ where $(z,w)\in \mathbb{R}^+\times\mathbb{R}^+$ for positive-positive feedback and $(z,w)\in \mathbb{R}^+\times\mathbb{R}^-$ for negative-negative feedback.
\end{itemize}
\end{proposition}
\begin{proof} Let $(a,b)$ be an equilibrium of (\ref{Reduced_model}), then 
$0 = -a + p\tanh(b)$ and $0 = -qb + r\tanh(a)$,
it follows that
\begin{equation}\label{b_values}
b= \operatorname{arctanh}(a/p)\quad\mbox{and}\quad b= r\tanh(a)/q,
\end{equation}
thus, $a$ is a solution to the transcendental equation
\begin{equation}\label{a_values}
\operatorname{arctanh}(a/p)= r\tanh(a)/q.
\end{equation}

\textbf{Case 1:} $p, r >0$ (positive-positive feedback). Let 
\begin{center}
    $h(x) = (r/q)\tanh(x)-\operatorname{arctanh}(x/p), \quad x \in (-p, p)$,
\end{center}
thus, the solutions of (\ref{a_values}) are the roots of $h$. Using Lemma~\ref{Lemma_h_function}, we have that if $q < pr$, then there exists $z\in\mathbb{R}^+$ such that $z,-z$ and $0$ are all possible roots of $h$, and if $q \geq pr$, 0 is the only possible root of $h$. Using (\ref{b_values}), we finish the proof.

\textbf{Case 2:} $p, r <0$ (negative-negative feedback). Since $\operatorname{arctanh}$ is an odd function and in this case $p=-|p|$, then (\ref{b_values}) becomes
\begin{equation}\label{b_values_neg_neg}
b= -\operatorname{arctanh}\left(\frac{a}{|p|}\right)\text{ and } b= -\frac{|r|}{q}\tanh(a),
\end{equation}
which is equivalent to (\ref{a_values}) replacing $p$ with $|p|$ and $r$ with $|r|$. Since $|p|, |r|>0$, using the same arguments as in the previous case, we finish the proof. 
\end{proof}

\begin{proposition}\label{equilibrium_mix}
If the system (\ref{Reduced_model}) is considered with positive-negative feedback, i.e. $pr<0$, then $\hat{0}$ is the only equilibrium point.
\end{proposition}
\begin{proof} Let $(a,b)$ be an equilibrium of (\ref{Reduced_model}) with positive-negative feedback. By (\ref{b_values}) and (\ref{a_values}) it follows that $\operatorname{arctanh}(a/p)= r\tanh(a)/q$.
If $p<0<r$ or $r<0<p$, each side of the equality has opposite signs, therefore it follows that $\hat{0}$ is the only possible equilibrium point.
\end{proof}

Next, we will show results concerning the linear stability of the equilibrium points reported in Propositions \ref{equilibrium_positive_and_negative_feedback} and \ref{equilibrium_mix}.

The Jacobian matrix of the system (\ref{Reduced_model}) is given by
\scalebox{0.85}{$J(x,y) = \begin{pmatrix}
-1 & p\mbox{sech}^2(y) \\
r\mbox{sech}^2(x) & -q 
\end{pmatrix}$},
therefore, its characteristic polynomial is 
\begin{equation}\label{characteristic_pol}
    \lambda^2 + (q+1)\lambda + q-pr\operatorname{sech}^2(y)\operatorname{sech}^2(x), 
\end{equation}
and the eigenvalues are given by 

\begin{equation}\label{Eigenvalues_pos_pos_neg_neg}
\footnotesize \lambda= \frac{-(q+1) \pm\sqrt{(q-1)^2+4pr\operatorname{sech}^2(y)\operatorname{sech}^2(x)}}{2} 
\end{equation}

\begin{theorem}\label{stability_positive_negative_feedback}
If the system (\ref{Reduced_model}) is considered with either positive-positive or negative-negative feedback, i.e. $pr >0$, then (\ref{Reduced_model}) has a supercritical pitchfork bifurcation such that
\begin{itemize}[leftmargin=0.7cm, itemsep = 0.05cm, topsep=0.05cm]
    \item[i)] $\hat{0}$ is stable, if $q> pr$,   
    \item[ii)] $\hat{0}$ is unstable, and   $(z,w)$, $-(z,w)$ are stable if $q< pr$, where $(z,w)\in \mathbb{R}^+\times\mathbb{R}^+$ for positive-positive feedback and $(z,w)\in \mathbb{R}^+\times\mathbb{R}^-$ for negative-negative feedback.
\end{itemize}
\end{theorem}

\begin{figure*}
\centering
\includegraphics[width=0.95\linewidth]{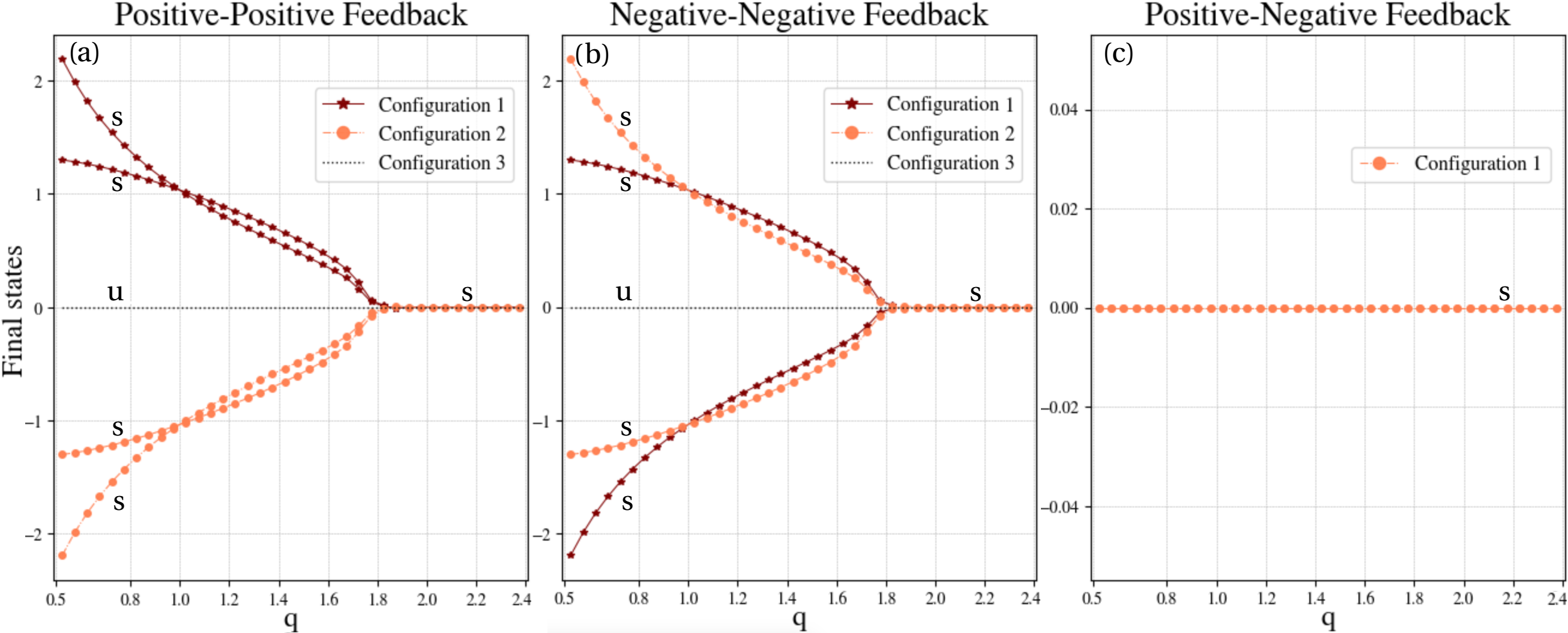}
\caption{\label{Figure_Bifurcation_diagram}Bifurcation diagrams of the system (\ref{Reduced_model}) with $|p|=|r|=4/3$ (particularly $pr\approx 1.77$). The system has three different equilibrium points when considering strong feedback (i.e. when $q<pr$), see (a) and (b) for positive-positive and negative-negative feedback, respectively. Note that the actors' equilibrium states (configurations) have the same sign with positive-positive feedback and opposite signs otherwise. When the feedback is positive-negative (see (c)) or $q> pr$, then $\hat{0}$ is the only equilibrium point (both values of each possible configuration of the system are equal to $0$). In these diagrams $s$ and $u$ denote stable and unstable entries of equilibrium points, respectively.}
\end{figure*}

\begin{proof} \textbf{Case 1:} $q<pr$. Let $\lambda_1$ be the eigenvalue with positive sign in (\ref{Eigenvalues_pos_pos_neg_neg}). Evaluating at the equilibrium point $\hat{0}$, we have 
\begin{center}
\small $\lambda_{1}   = \left(-(q+1) +\sqrt{(q+1)^2-4(q-pr)}\right)/2
     >\left(-(q+1) + (q+1)\right)/2=0.$
\end{center}
Thus, $\hat{0}$ is unstable.  

\textbf{Subcase 1:} $p,r>0$. Let $f$ be as in Lemma \ref{Lemma_h_function}, Thus, by Case 1 of the Lemma, there exists $u$ in the interval $(0,z)$ such that $f(u)=q/(pr)$ see Fig.~\ref{h_and_f}. Since $f$ is decreasing, then 
    $f(z)<f(u)= q/(pr)$. 
By (\ref{b_values}) it follows that $f(z)= \operatorname{sech}^2(w)\operatorname{sech}^2(z)$, therefore 
\begin{equation}\label{Stability_inequality}
0<q-pr\operatorname{sech}^2(w)\operatorname{sech}^2(z).
\end{equation}
Thus, all coefficients of (\ref{characteristic_pol}) are positive, so by the Routh-Hurwitz Criterion we conclude that $(z,w)$ is stable. Trivially, (\ref{Stability_inequality}) is preserved if we change $z$ to $-z$ and $w$ to $-w$, so $-(z,w)$ is stable (see Fig. ~\ref{Figure_Bifurcation_diagram} (a) for an illustration of this).

\textbf{Subcase 2:} $p, r<0$. Exchanging $p$ with $|p|$, $r$ with $|r|$ and using (\ref{b_values_neg_neg}), it follows that this subcase is reduced to Subcase 1, consequently we also conclude that $(z,w)$ and $-(z,w)$ are stable (see Fig.~\ref{Figure_Bifurcation_diagram} (b) for an illustration of this).

\textbf{Case 2:} $q> pr$. In this case, $\hat{0}$ is the only equilibrium point of (\ref{Reduced_model}), so (\ref{characteristic_pol}) becomes  $\lambda^2 + (q+1)\lambda + q-pr$. Since $q+1>0$ and $q-pr\geq 0$, by the Routh-Hurwitz Criterion we conclude that $\hat{0}$ is stable.
\end{proof}
\begin{theorem}\label{0_onlyEquilibrium__mixed_feedback}
If the system (\ref{Reduced_model}) is considered with positive-negative feedback, i.e.
$pr < 0$, then the only equilibrium point, $\hat{0}$, is stable.
\end{theorem}

\begin{proof} In Proposition \ref{equilibrium_mix} we proved that $\hat{0}$ is the only equilibrium point of (\ref{Reduced_model}), so by evaluating (\ref{characteristic_pol}) at $\hat{0}$ we get the characteristic polynomial
    $\lambda^2 + (q+1)\lambda + q-pr$. 
Then, since all coefficients are positive when $pr< 0$, by the Routh-Hurwitz Criterion  $\hat{0}$ is stable (Fig.~\ref{Figure_Bifurcation_diagram} (c)).
\end{proof}

\begin{figure*}
\centering
\includegraphics[width=1\linewidth]{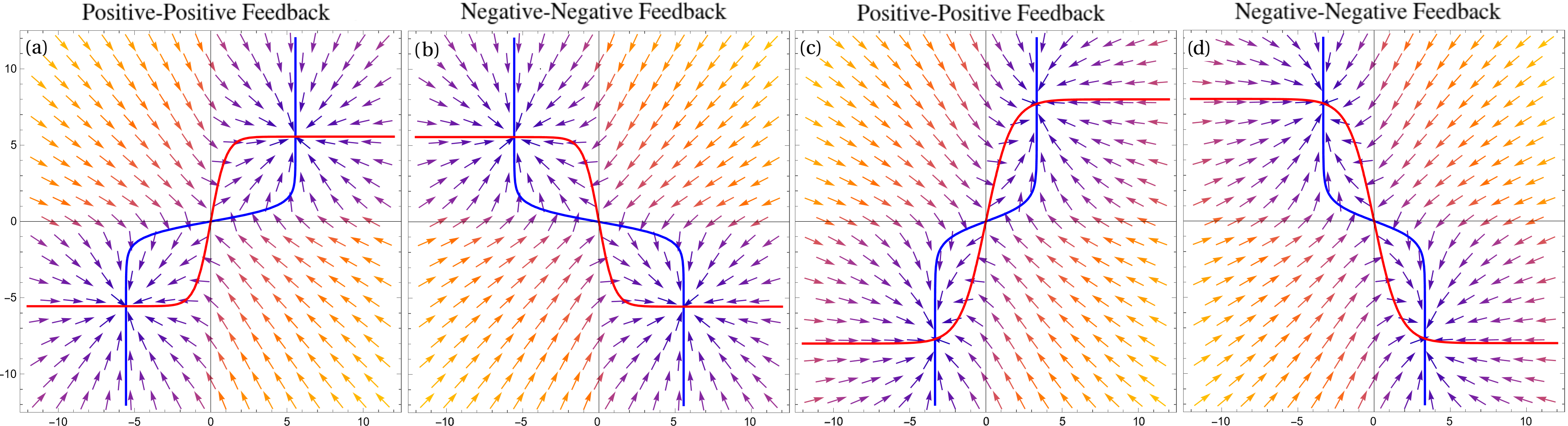}
\caption{\label{Figure_Phase_portrait}In (a) and (b) the phase portraits of the system (\ref{neg_2actors}) for strong positive-positive and strong negative-negative feedback, respectively. In (c) and (d) the phase portraits of the system (\ref{NEW_2_actors_model}) for strong positive-positive and strong negative-negative feedback, respectively. For (a) and (b) parameter values were set at $m_1=m_2=0.9, |c_1|=|c_2|=5$; and $m_1=0.9, m_2=0.5, |c_1|= 3, |c_2|=4, p_1= 0.8, p_2 = 0.6$ for (c) and (d). Note the nullcline sets (solid lines) and the pockets (closed regions where the direction of flow is oriented only towards a unique direction) have the same topological configurations for systems (\ref{neg_2actors}) and (\ref{NEW_2_actors_model}).}
\end{figure*}

\begin{remark}\label{Remark_P_condition}
In general, the bifurcation of the system (\ref{Reduced_model}) with strong feedback occurs when $q<pr$, as seen in Fig.~\ref{Figure_Bifurcation_diagram}, if and only if $p>1$. Consider $p\leq 1$; then, the vertical nullcline (blue curve in Fig.~\ref{Figure_Phase_portrait} (c)) always remains above the identity line, thus the intersection of both nullclines can never occur on the identity line, i.e. the equilibrium states of both actors cannot be equal. Analytically, one may verify that, defining $f(x)= x-\operatorname{arctanh}(x/p)$, then 
    $f(x)=0, \ x\in(0,\infty)$ iff $p>1$,
which follows from Lemma \ref{Lemma_h_function}.
\end{remark}

\begin{theorem}\label{Limit_cycles}
The system (\ref{NEW_2_actors_model}) does not have limit cycles.
\end{theorem}

\begin{proof}  Theorem 1 in \onlinecite{Liebo} proves that system (\ref{neg_2actors}) does not have
limit cycles.  We will use similar arguments  to prove the absence of limit cycles in our model.

When the system (\ref{NEW_2_actors_model}) is considered with either positive-positive or negative-negative feedback, the nullclines and pockets (closed regions where the flow is oriented in a single direction) have the same topological configuration as the respective nullcline sets and pockets of the system (\ref{neg_2actors}) since $m_i, p_i>0, i=1,2$. Fig.~\ref{Figure_Phase_portrait} shows the situation for strong feedback. Thus, Theorem 1 in \onlinecite{Liebo} applies, and no limit cycles exist for system (\ref{NEW_2_actors_model}) in these cases.

On the other hand, suppose that the system (\ref{NEW_2_actors_model}) has positive-negative feedback, i.e. $c_1c_2<0$. Without loss of generality, assume that $c_2<0$ and define $H:\mathbb{R}^2\rightarrow \mathbb{R}$, 
$H(x_1,x_2)= -c_2\int_0^{x_1}\tanh(p_2u)du + c_1\int_0^{x_2}\tanh(p_1u)du.$
Since 
\begin{eqnarray*}
\dot{H}&=& -c_2\tanh(p_2x_1)(-m_1x_1+c_1\tanh(p_1x_2)\\ 
&&+ c_1\tanh(p_1x_2)(-m_2x_2+c_2\tanh(p_2x_1)\\
&=&c_2\tanh(p_2x_1)m_1x_1- c_1\tanh(p_1x_2)m_2x_2,
\end{eqnarray*}
then $H$ is a Lyapunov function, therefore the system has no limit cycles.
\end{proof}

\begin{theorem}
    \label{manifolds}
    The stable and unstable manifolds of the saddle point at the origin of the system (\ref{Reduced_model}), when it exists, are odd functions when expressed as curves in phase space, $y=f(x)$. Furthermore, when $q=1$ and $r=p$, these are straight lines with slope one and minus one.
\end{theorem}
\begin{proof}
Trivially, the system is invariant under the change of variables $x=-u$, $y=-v$. Since the saddle point lies at the origin, the stable and unstable manifolds as curves in phase space, $y=f(x)$ must contain the origin, thus $f(0)=0$. 
Furthermore, from the aforementioned invariance, we have $-v=f(-u)$. Thus, the manifolds are invariant only if $f(x)$ is odd. Finally, from the chain rule, we know that ${dy\over dx}=(\frac{dy}{d\tau})/(\frac{dx}{d\tau}).$ From the system (\ref{Reduced_model}) we obtain that $    {dy\over dx}=\frac{-qy+r\tanh(x)}{-x+p\tanh(y)}.$
 For $y=f(x)$ to be a straight line, necessarily $dy/dx=m$, with $m\in\mathbb{R}$ some constant. We see that this is achieved when $y=x$ or $y=-x$, and $\frac{r}{q}=p$. Furthermore, such that both conditions $y=x$ (or $y=-x$) and $dy/dx=m$ are fulfilled simultaneously, $q=1$ so $r=p$ follows.
 \end{proof}

\section{Discretization of the model}
To discretize the system of differential equations of the system (\ref{Reduced_model}), in this paper we use a nonstandard finite-difference scheme, developed in \onlinecite{mickens1994nonstandard}. These allow to approximate a system of differential equations through a system of difference equations
consistently, preserving stability properties of the equilibrium points, a finite-difference scheme with this property is called elementary stable (see Definition 11 in \onlinecite{anguelov2001contributions}). Specifically, we use the nonstandard forward Euler finite-difference scheme given in Theorem 13 in \onlinecite{anguelov2001contributions} for the cases of positive-positive and negative-negative feedback, and Theorem 1 in \onlinecite{dimitrov2005nonstandard}  for the case of positive-negative feedback.
\begin{lemma}\label{Justificacion_hipotesisDiscretizacion1}
If the system (\ref{Reduced_model}) is considered with either positive-positive or negative-negative feedback, i.e. $pr >0$, then (\ref{Reduced_model}) satisfies the hypotheses of the Theorem 13 in \onlinecite{anguelov2001contributions}  with 
$M:=\max\{|\lambda|: \lambda\in \Omega\}=|\lambda_2|$,
where $\lambda_2$ is the eigenvalue associated with the equilibrium point $\hat{0}$, with negative sign in (\ref{Eigenvalues_pos_pos_neg_neg}) and $\Omega=\bigcup_{\hat{y}\in\Gamma}\sigma(J(\hat{y}))$ with $\Gamma$ the set of all equilibrium points of the system (\ref{Reduced_model}) and $\sigma(J(\hat{y}))$ the set of eigenvalues of the Jacobian of the system (\ref{Reduced_model}) at $\hat{y}$.
\end{lemma}
\begin{proof}  By Proposition \ref{equilibrium_positive_and_negative_feedback},
if he system (\ref{Reduced_model}) is considered with either positive-positive or negative-negative feedback, then it has either one or three equilibrium points. Also, note that the expression inside the root in (\ref{Eigenvalues_pos_pos_neg_neg}) satisfies
\begin{equation}\label{Inside_root_1}
(q-1)^2+4pr\operatorname{sech}^2(y)\operatorname{sech}^2(x)>0, 
\end{equation}
since $pr>0$; therefore, all eigenvalues are real.

Let $\lambda_1$, $\lambda_2$, be the eigenvalues corresponding to $\hat{0}$ with positive and negative signs in (\ref{Eigenvalues_pos_pos_neg_neg}), respectively.

\textbf{Case 1:} $q>pr$.  As shown in Proposition \ref{equilibrium_positive_and_negative_feedback},
 $\hat{0}$ is the only equilibrium point. Using Theorem \ref{stability_positive_negative_feedback}, we conclude that the roots of the characteristic polynomial (\ref{characteristic_pol}) are real and negative.

Define 
$$
\theta= (q+1)/2\quad\mbox{y}\quad\gamma = \sqrt{(q-1)^2+4pr}/2.$$
Then by the definition of $q$ and (\ref{Inside_root_1}) it follows that $\theta, \gamma$ are positive reals. Furthermore, in this case, it follows that
$$-\theta-\gamma=\lambda_2<\lambda_1=-\theta+\gamma<0,$$
so $|\lambda_1|<|\lambda_2|.$ 
Therefore, we conclude that 
$M=|\lambda_2|.$

\textbf{Case 2:} $q< pr$. Given $\theta$ and $\gamma$ as in Case 1, here
$$-\theta-\gamma=\lambda_2<0<\lambda_1=-\theta+\gamma,$$
holds, where the second inequality was proven  in Theorem \ref{stability_positive_negative_feedback}. Therefore, 
    $|\lambda_2|=\theta+\gamma> -\theta+\gamma = |\lambda_1|.$ 

Let $\lambda_3$, $\lambda_4$, be the eigenvalues associated with the equilibrium point $(z,w)$ different from $\hat{0}$, with positive and negative signs in (\ref{Eigenvalues_pos_pos_neg_neg}), respectively and $\lambda_3'$, $\lambda_4'$, the eigenvalues associated with the equilibrium point  $-(z,w)$ taking positive and negative signs in (\ref{Eigenvalues_pos_pos_neg_neg}), respectively. Since  $\operatorname{sech}$ is an even function, then 
    $\lambda_3= \lambda_3'\quad \mbox{and} \quad\lambda_4= \lambda_4'.$
Using Theorem  \ref{stability_positive_negative_feedback}, we conclude that the roots of the characteristic polynomial (\ref{characteristic_pol}) are real and non-zero for each equilibrium point.

Let 
\scalebox{0.95}{$\alpha= \sqrt{(q-1)^2+4pr\operatorname{sech}^2(y)\operatorname{sech}^2(x)}/2.$}
Using (\ref{Inside_root_1}), it follows that $\alpha>0$. Thus, by arguments similar to those used previously, it follows that $|\lambda_3| <|\lambda_4| = \theta+\alpha.$ Since $\operatorname{sech}(x)<1$ and  $\alpha<\gamma$, thus
    $|\lambda_4|<|\lambda_2|.$ 
Therefore, $M = |\lambda_2|.$\end{proof}

\begin{lemma}\label{Justificacion_hipotesisDiscretizacion2}
Considering the system (\ref{Reduced_model})  with positive-negative feedback, i.e. $pr <0$,  satisfies the hypotheses of Theorem 1 in \onlinecite{dimitrov2005nonstandard} with
$M:=\max\{|\lambda|^2/( 2|\operatorname{Re}(\lambda)|): \lambda\in \Omega\}= |\lambda_2|^2/( 2|\operatorname{Re}(\lambda_2)|),$
where $\lambda_2$ and $\Omega$ are as in Lemma \ref{Justificacion_hipotesisDiscretizacion1}.
\end{lemma}

\begin{proof} According to Proposition \ref{equilibrium_mix}, $\hat{0}$ is the only equilibrium point with positive-negative feedback. Following Theorem \ref{0_onlyEquilibrium__mixed_feedback},  $\hat{0}$ is stable, i.e. the roots of (\ref{characteristic_pol}) have negative real parts when $pr<0$. 

Let $\lambda_1, \lambda_2, \theta, \gamma$, as in Lemma \ref{Justificacion_hipotesisDiscretizacion1},
\begin{equation}\label{Lambda_theta_gamma}
\lambda_2 = -\theta-\gamma,\quad\lambda_1=-\theta+\gamma.
\end{equation}
Note that in this case $\gamma$ can be either real or complex.

\textbf{Case 1:} $\gamma\in\mathbb{R}$. 
In this case, $\lambda_2<\lambda_1<0$,
or equivalently, $|\lambda_1|<|\lambda_2|$. It follows that $M={|\lambda_2|^2\over 2|\operatorname{Re}(\lambda_2)|}.$

\textbf{Case 2:} $\gamma\in\mathbb{C}\setminus\mathbb{R}$. In this case, by (\ref{Lambda_theta_gamma}), 
$|\lambda_1|=|\lambda_2|$ and $Re(\lambda_1) = Re(\lambda_2),$ therefore, 
$M={|\lambda_2|^2\over 2|\operatorname{Re}(\lambda_2)|}.$
Furthermore, in this case 
$M=(q-pr)/(q+1)$
holds.
\end{proof}

\begin{theorem}
\label{Discretizacion_final_Modelo_OD}
The system (\ref{Reduced_model}) has an elementary stable scheme of difference equations of the form
\begin{equation}\label{Nonstandard_scheme_Euler}
\begin{array}{lcl}
x(n+1)&=& Cx(n)+P\tanh(y(n))\\
y(n+1)&=& Qy(n)+R\tanh(x(n)),
\end{array}
\end{equation}
where $n\in\mathbb{N}$ and $C, Q, P, R\in \mathbb{R}$ are such that  
\begin{itemize}[leftmargin=0.7cm, itemsep = 0.05cm, topsep=0.05cm]
    \item[i)] $Q, P, R>0$ for positive-positive feedback,   
    \item[ii)] $Q>0$ and  $P, R<0$ for negative-negative feedback,
    \item[ii)] $PR<0$ for positive-negative feedback.
\end{itemize}
Furthermore, if the step size $h$ satifies $0<h\leq 1$, then $C\in\mathbb{R}^+$.
\end{theorem}
\begin{proof} Let $q,p,r$ as in the system (\ref{Reduced_model}). By Lemmas \ref{Justificacion_hipotesisDiscretizacion1} and  \ref{Justificacion_hipotesisDiscretizacion2}, it follows that (\ref{Reduced_model}) satisfies the hypotheses of Theorem 13 in \onlinecite{anguelov2001contributions} and Theorem 1 in \onlinecite{dimitrov2005nonstandard}, respectively. So if we take  $\phi(h) = 1 - e^{-h}$ in both cases (see Remark 14 in \onlinecite{anguelov2001contributions}), then (\ref{Reduced_model}) has an elementary nonstandard
forward Euler finite-difference scheme of the form
\begin{center}
\small$\begin{array}{lcl}
x(n+1)&=& x(n)+\dfrac{\phi(hM)}{M}(-x(n)+p\tanh(y(n))\\

y(n+1)&=& y(n)+\dfrac{\phi(hM)}{M}(-qy(n)+r\tanh(x(n)), 
\end{array}$
\end{center}
with $M$ as in Lemmas \ref{Justificacion_hipotesisDiscretizacion1} and  \ref{Justificacion_hipotesisDiscretizacion2}. Let  
\begin{eqnarray}\label{Difference_parameters}
&& C= 1- (\phi(hM)/M),\quad Q= 1- (q\phi(hM)/M),\nonumber \\
&&P= (p\phi(hM))/M, \quad R=(r\phi(hM))/M,
\end{eqnarray}
since $\phi(hM)/M>0$, it follows that
$\operatorname{sgn}(P)=\operatorname{sgn}(p)$ and $\operatorname{sgn}(R)=\operatorname{sgn}(r)$. 
    
If $pr>0$, i.e., if the system (\ref{Reduced_model}) has positive-positive or negative-negative feedback, by (\ref{Inside_root_1}), 
\begin{equation}
\small    \label{M_definition}
q+1+\sqrt{(q-1)^2+4pr}>q+1+(q-1)=2q,
\end{equation}
then $M>q$. Since $0<\phi(z)<1$ for $z>0$, thus $M>q\phi(hM)$, so $Q>0$.

Finally, suppose $0<h\leq 1$. Let $f:[0,\infty)\rightarrow\mathbb{R}$,
    $f(M)= M-\phi(hM)$.
Note that $f'(M)= 1-h/e^{Mh}>0$, since $e^{Mh}>h$ for $0<h\leq 1$. Therefore, $f$ is increasing, and since $f(0)=0$, it follows that $f(M)>0$; therefore $C>0$. 
\end{proof}

\begin{theorem}
\label{Theorem_Discrete_statements}
The following statements are satisfied for the elementary stable scheme of difference equations (\ref{Nonstandard_scheme_Euler}):
\begin{itemize}[leftmargin=0.5cm, itemsep = 0.05cm, topsep=0.05cm]
    \item[] For positive-positive feedback
        \begin{itemize}
        \item[a)] if $\operatorname{sgn}(x(0))= \operatorname{sgn}(y(0))$, then $\operatorname{sgn}(x(0))=\operatorname{sgn}(x(n))= \operatorname{sgn}(y(n))$,   $n\in\mathbb{N}$,
        \item[b)] when $q=1$, $p=r$ and   $\operatorname{sgn}(x(0))= \operatorname{sgn}(y(0))$, it follows that if $y(0)<x(0)$, then $y(n)< x(n)$,  $n\in\mathbb{N}$,
        \item[c)] when $q=1$, $p=r$ and   $\operatorname{sgn}(x(0))= \operatorname{sgn}(y(0))$, it follows that $\{x(n)\}_{n\in\mathbb{N}}$ and $\{y(n)\}_{n\in\mathbb{N}}$ converge to the same value $\xi$ such that $\operatorname{sgn}(\xi)=\operatorname{sgn}(x(0))$.
        \end{itemize}
    \item[] For negative-negative feedback 
    \begin{itemize}
        \item[d)] if $y(0)<0< x(0)$, then $y(n)<0< x(n)$, $n\in\mathbb{N}$,
        \item[e)] when $q = 1$, $p = r$ and $\operatorname{sgn}(x(0))= -\operatorname{sgn}(y(0))$, if $|y(0)|<x(0)$, then $|y(n)|<x(n)$,  $n\in\mathbb{N}$,
        \item[f)] when $q = 1$, $p = r$ and $y(0)<0< x(0)$, then there exists $\xi\in\mathbb{R}^+$ such that  $\{y(n)\}_{n\in\mathbb{N}}$ and $\{x(n)\}_{n\in\mathbb{N}}$ converge to  $-\xi$ and  $\xi$, respectively.
    \end{itemize}
\end{itemize}
\end{theorem}
\begin{proof} a) and d) are immediate by induction. 

To prove b), let observe that, by (\ref{Difference_parameters}), it follows that $Q=C$ and $P=R$. Suppose that $y(0)<x(0)$, then
\begin{eqnarray*}
\small y(1)<x(1) &\iff & Cy(0)+P\tanh(x(0))\\
&&< Cx(0)+P\tanh(y(0)) \\
     &\iff & Cy(0)-P\tanh(y(0))\\
&&< Cx(0)-P\tanh(x(0)). 
\end{eqnarray*}
Define $h(x)=Cx-P\tanh(x)$, we must show that $h$ is increasing. Since $h'(x)=C-P\operatorname{sech}^2(x),$
we must show that $C/P>\operatorname{sech}^2(x)$. Substituting $q=1$ and $p=r$ in (\ref{M_definition}), it follows that $M=p+1$, since $0<\phi(z)<1$, for $z>0$. Note that $1>{\phi(hM)(p+1)\over M},$
which is equivalent to
$C/P>1$. Since  $1\geq \operatorname{sech}^2(x)$, $x\in\mathbb{R}$, then $C/P>\operatorname{sech}^2(x)$. 

To prove e), observe that, if $|y(0)|<x(0)$, then 
\begin{eqnarray*}
|y(1)|<x(1) \iff C|y(0)|-|P|\tanh(|y(0)|)\\
< Cx(0)-|P|\tanh(x(0)), 
\end{eqnarray*}
thus, this case is analogous to b).

To prove c), assume, without loss of generality, $0<y(0)<x(0)$. Then by b) it follows that 
\begin{eqnarray}\label{for_step_diagram}
Cy(n)+P\tanh(y(n))&<& x(n+1), y(n+1)\nonumber\\
    <Cx(n)&+&P\tanh(x(n)).
\end{eqnarray}
Let $g(x)=Cx+P\tanh(x)$, then there exists $\xi\in\mathbb{R}^+$ such that $\xi$ is the only positive fixed point of $g$. This follows from $C/P>1$, which was shown for b). Thus, any sequence obtained by iterating $g$ starting from a positive initial value always converges to $\xi$. Therefore, the extremum points of (\ref{for_step_diagram}) converge to $\xi$, as well as$\{x(n)\}_{n\in\mathbb{N}}$ and
$\{y(n)\}_{n\in\mathbb{N}}$. 

Finally, f) is analogous to c), since if $|y(0)|<x(0)$, it follows that 
\begin{center}
\small $C|y(n)|+|P|\tanh(|y(n)|)< x(n+1)= Cx(n)
    +|P|\tanh(|y(n)|)<Cx(n)+|P|\tanh(x(n)),$
\end{center}
\begin{center}
\small $-\displaystyle[ Cx(n)+|P|\tanh(x(n))]< y(n+1)= 
    -\displaystyle[ C|y(n)|+|P|\tanh(x(n))]<-\displaystyle[C|y(n)|+|P|\tanh(|y(n)|].$
\end{center}
\end{proof}

\end{document}